\documentstyle[12pt,epsfig]{article}
\input epsf.tex
\bf
\begin{document}
\title{On the muon neutrino mass}
\date{}
\maketitle              
\author{N. Angelov$^{2}$, \and F. Balestra$^{1}$, \and Yu. Batusov$^{2}$,
 \and A. Bianconi$^{3}$, \and
M.P. Bussa$^{1}$,
 \and L. Busso$^{1}$, \and L. Ferrero$^{1}$,
 \and R. Garfagnini$^{1}$, \and I. Gnesi$^{1,5}$, \and
E. Lodi Rizzini$^{3}$,  \and A. Maggiora$^{1}$,
 \and D. Panzieri$^{4}$,  \and G. Piragino$^{1,5}$,
 \and G. Pontecorvo$^{2,5}$,
 \and F. Tosello$^{1}$, and  \and L. Venturelli$^{3}$}
\begin{center}
%\smallsize
%
\footnotesize
$^{1}Dipartimento$ $di$ $Fisica$ $ Generale$ $''Amedeo$  $Avogadro'',$\\
$University$ $ of$ $Torino$; $INFN$, $Sez.$ $di$ $Torino$,
$ Torino, Italy$\\
$^{2}Joint$ $Institute$ $for$ $Nuclear$ $Research$, $ Dubna$, $Russia$\\
$^{3}Dipartimento$ $di$ $Chimica$ $e$ $Fisica$ $per$ $l'Ingegneria$ $e$
 $per$ $i$ $Materiali$,\\
$University$ $of$ $Brescia$; $INFN$, $Gruppo$ $collegato$ $di$ $Brescia$,
$Brescia$, $Italy$\\
$^{4}Dipartimento$ $di$ $Scienze$ $e$ $Tecnologie$ $Avanzate$,\\
$University$ $of$ $Piemonte$ $Orientale$; $INFN,$ $Gruppo$ $collegato$ $di$ $Alessandria,$\\
$Alessandria$, $Italy$\\
$^{5}Centro$ $Studi$ $e$ $Ricerche$ $''Enrico$  $Fermi''$, $Roma$, $Italy$
\normalsize
\end{center}
\begin{abstract}
During the runs of the PS 179 experiment at LEAR of CERN,
we photographed an event of antiproton-$Ne$ absorption, with a
complete $\pi^{+}$$\rightarrow$$\mu^{+}$$\rightarrow$e$^{+}$
chain.
From the vertex of the reaction a
very slow energy $\pi^{+}$ was emitted. The $\pi^{+}$ decays
into a $\mu^{+}$ and
subsequently the $\mu^{+}$ decays into a positron. At the first decay vertex
a muon neutrino was emitted and at the second decay vertex an
electron neutrino and a muon antineutrino. Measuring the pion
and muon
tracks and applying the momentum and energy conservation and
using a classical
statistical interval estimator, we obtained an experimental 
upper limit for the muon neutrino mass: $m_{\nu}$ $<$ 2.2 MeV
at a 90$\%$ confidence level. 
A statistical analysis has been performed
of the factors contributing to the square value
of the neutrino mass limit.

PACS. 13.20 Cz, 14.60 Pq, 13.25 Cq  
%\vspace{1.0cm}
%\begin{center}
%(EPJ - accepted for publication)
%\end{center}

\end{abstract}

\newpage

\section{Introduction}

Determination of the absolute values of 
neutrino masses represents a most difficult problem from an
experimental point of view. Evidence in favour of non-zero neutrino masses
and oscillations obtained in most of the relevant neutrino 
experiments has made the physics of massive neutrinos
a frontier field of research in particle physics and astrophysics. 
All the existing terrestrial and astrophysical data indicate
that the neutrino masses are by many orders of magnitude 
smaller than those of other
experimentally measured lepton and hadron masses. 
Such a low value is the most relevant
reason for it being extremely difficult to extract
the values of neutrino masses
from experimental measurements, either directly or indirectly. 
This is the main reason why most 
experimental papers only
report a confidence limit interval for such values. 
In this paper we report an upper limit
for the muon neutrino
mass obtained by measuring the radii of curvature of the pion and
muon tracks of
an event, in which a pion decays into a muon and a muon neutrino, 
recorded during the runs of the PS 179 experiment at the
beam of antiprotons of LEAR at CERN.

\section{Experimental apparatus}

\begin{figure}
\begin{center}
\epsfig{file=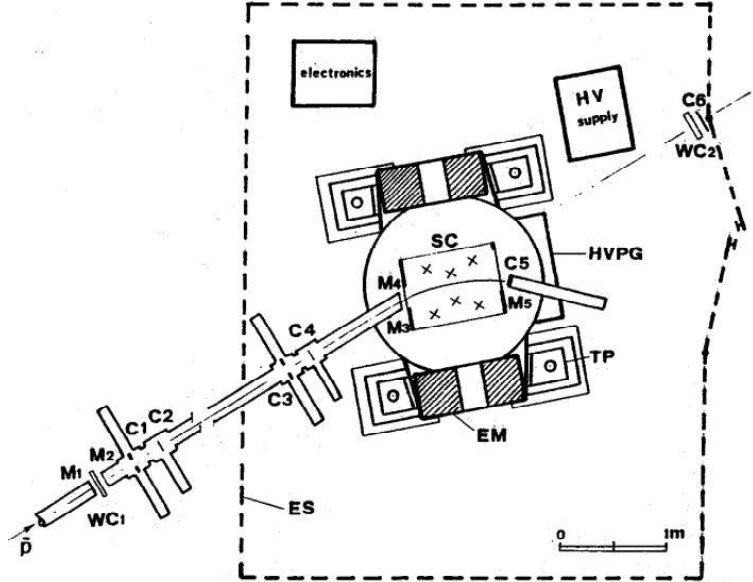,width=12cm}
\caption{Layout of the PS 179 experimental apparatus, including:
 EM-electromagnet; 
SC-streamer chamber; HVPG-high voltage generator; TP-travelling platform;
ES-electrostatic screening; $WC_{1-2}$-wire chambers; $C_{1-6}$
-scintillation counters; $M_{1-5}$-thin walls.}
\label{Figure 1.}
\end{center}
\end{figure}

The PS 179 experimental apparatus was designed and
built for the study of antiproton interactions with
light and medium-light nuclei at the LEAR facility of CERN. The
aim of the research was an experimental study
of the interaction of antiprotons with nuclei at low energies.
The results obtained provided information 
on the fundamental nucleon-antinucleon forces,
on the interaction of antiprotons with clusters of bound nucleons, 
on the distribution of nuclear matter, and on properties of highly 
excited nuclear matter, as well as a restriction on the possible amount 
of antimatter present in the early Universe~\cite{PS179}.
In order to make the most of the information available on all the
secondary charged particles produced in the reactions, 
in designing the experiment a choice was made of
the visualization detection technique, and 
a self-shunted streamer chamber placed in a magnetic field was used.
Such a detector had many  advantages. It was
a low density gas target, and, at the same time, 
it was triggerable, offering a 
4$\pi$ acceptance in which highly luminous localized 
particle tracks could be obtained ~\cite{NIM75}. The low density 
of the target medium, i.e. of the streamer chamber filling gas 
at 1 atm pressure, 
allowed to reveal long charged particle ranges and nuclear fragments. 
The experimental apparatus used is sketched in Figure~\ref{Figure 1.}.
The detector provided stereo pictures of the sensitive chamber volume.
The details of the experimental apparatus have been given  
in ref.~\cite{NIM85} where a more complete description of the 
characteristics and the performances of all the components
of the setup can be found. 

\section{Analysis of the measurements}

The runs of the PS 179 experiment were carried out with the
streamer chamber filled at 1 atm with different gaseous targets: $^{3}He$,
$^{4}He$, $Ne$. The stereo pictures of events, recorded 
on films during 
the runs, were reprojected onto measuring tables for visual 
scanning. The events were measured with a 
digitized coordinatometer directly on the scanning tables.

\begin{figure}
\begin{center}
\epsfig{file=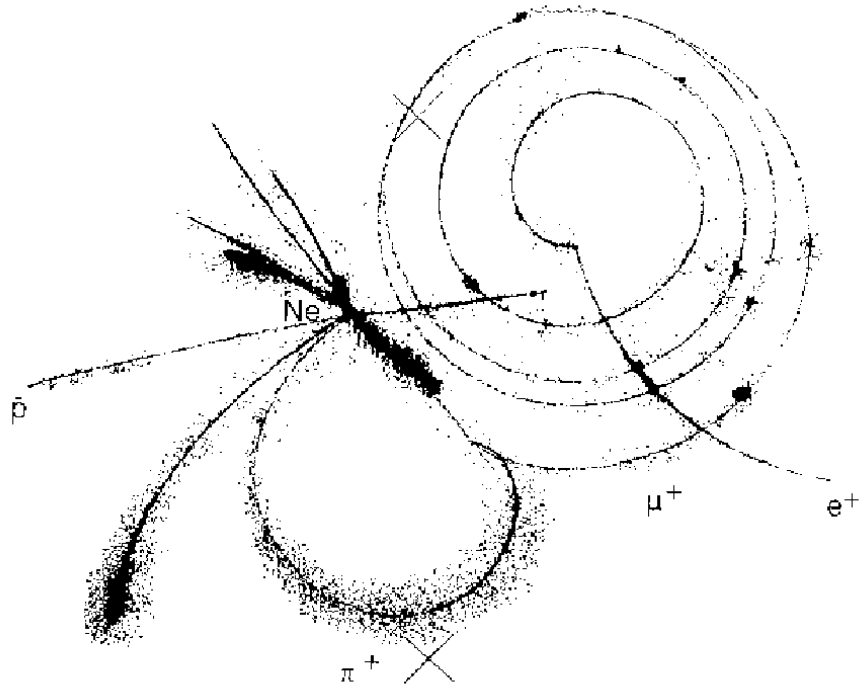,width=12cm}
\caption{PS 179 picture of ``The Three Neutrino Event''.
An antiproton annihilates with the $Ne$ nucleus. 
From the interaction vertex a $\pi^{+}$ of energy
inferior to 2.0 MeV is emitted. The $\pi^{+}$ decays 
into a $\mu^{+}$, and 
subsequently the $\mu^{+}$ decays into a positron. 
At the first decay vertex,
a muon neutrino is, also, emitted, and at the second decay vertex an 
electron neutrino and a muon antineutrino.}
\label{Figure 2.}
\end{center}
\end{figure}

In one of the photographs of the $Ne$ 
run, the beautiful event, reproduced in 
Figure~\ref{Figure 2.}, was observed. 
It represents a multi-nucleon annihilation~\cite{NP491}
of an antiproton with a $Ne$ nucleus 
(see, also, Figure 11 of ref.~\cite{NIM85}). 
From the vertex of the interaction there ``evaporated'' a (1.98$\pm$0.02) MeV
positive pion. From the two vertices 
of the pion and muon
decays, three neutrinos were emitted: from the first
a muon neutrino and from the second an electron neutrino and a
muon antineutrino.
The peculiarity of the event consists in the noticeable length 
of the $\pi^{+}$ and $\mu^{+}$ tracks and in their coplanarity.

The temperature of the streamer chamber filling gas
was 289 K, the pressure atmospheric, and the target density 
(0.80$\pm$0.01) mg/cm$^{3}$. 
The magnetic field was 0.8 T ($\Delta$B/B = 10$^{-5}$). 

The tracks have been 
measured at JINR with a microscope-digitizer.
The recent acquisition of such a microscope made it possible to newly 
measure the $\pi\mu$e event, reconstructing and digitizing
tracks with a very high accuracy. 
The pion track has been digitized 
at 347 points, the muon track at 2037 points and the positron track at 
180 points. All the measured  points are approximately
 equidistant. Each point has 
been measured with a precision of about $10^{-2}$ mm. The coordinates of all 
track points have been used to deduce the radii of curvature and
the ranges. Each 
particle transfers energy by ionization
to the surrounding medium and slows down along its path. 
A Fortran code has been written to estimate  
the radii of curvature varying along the tracks and the total
track lengths. 
For each track
the best-fit circles were calculated taking into account
a fixed number N of points.
Starting from the first measured point and taking 
the subsequent N-1 points,
the first radius was calculated. If n is the total number of points measured 
along a track, shifting the N points by one along the track, 
the subsequent n-(N-1)
radii have been calculated. We have chosen N = 25 (a value
statistically significant in order to ensure fit stability)  
for each track. 
Program CIRCLE of the 
CERN library~\cite{CERN} has been included into the code to calculate 
the best-fit radii of curvature along the 
tracks.

The initial pion momentum was estimated from all measured 
points along the first
10 cm of its track, since the range-energy relations (and tables)
for neon at NTP reveal the energy loss of a 2.00 MeV pion
to be negligible for path lengths not exceeding 10 cm.
The average radius of curvature of the first 10 cm of the pion track,
obtained best-fitting the track with a circle,
is
\mbox{r = (9.79$\pm$0.09) cm} corresponding  to a momentum 
(p = 300Br)
of (23.50$\pm$0.15) MeV/c. The agreement between the value of the 
initial pion momentum and its measured path range of (471.5$\pm$0.8) mm,
shows that the pion decay occurred with a momentum
of (0.05$\pm$0.10) MeV/c.
The same parameters have been
measured for the muon track. 
The initial muon momentum was estimated from all measured points
along the first 12 cm of its track,
since, also in this case, the range-energy relations (and tables)
for neon at NTP reveal the energy loss of a 5.00 MeV muon
to be negligible for path lengths not exceeding 12 cm.
The value of the average radius of curvature 
of the first 12 cm of the
muon track, obtained with a best-fit circle, 
is \mbox{r = (12.50$\pm$0.08) cm}, corresponding to 
a momentum of (29.90$\pm$0.15) MeV/c. In this case, also,
the agreement between the initial muon momentum
and its measured path range of (2616.0$\pm$0.8) mm shows that
the muon decayed with a momentum of
(0.6$\pm$0.6) MeV/c.
The positron was emitted with a momentum of (46.80$\pm$0.08)
 MeV/c. 
The angle between the tangent lines of the pion and muon
trajectories at the decay vertex was (163.0$\pm$1.0) deg, 
and the angle between the muon and positron
tracks was (107.0$\pm$1.0) deg.
It should be pointed out that the $\pi\mu$e event of
Figure~\ref{Figure 2.} is the only existing of this kind whose parameters
could be all measured with high precision.

\subsection{Kinematic and dynamic constraints}

For kinematic and dynamic analysis of the reaction, the
relations of conservation of momentum and energy have been used.
The pion decay occurs in a plane. At the decay vertex,
the muon and the neutrino have flight directions at angles
 $\phi$ and $\theta$ to the pion flight
direction, respectively. For momentum conservation
\begin{equation}
 p_{\pi}=p_{\mu}cos\phi+p_{\nu}cos\theta
\end{equation}
\begin{equation}
 0=p_{\mu}sin\phi+p_{\nu}sin\theta      .
\end{equation}
For energy conservation 
\begin{equation}
E_{\pi}=E_{\mu}+E_{\nu}       .
\end{equation}
Combining algebraically the three relations one obtains
\begin{equation}
p_{\nu}^{2}=p_{\pi}^{2}+p_{\mu}^{2}-2p_{\pi}p_{\mu}cos\phi
\end{equation}
and
\begin{eqnarray}
E_{\pi}-E_{\mu}=\sqrt{p_{\pi}^{2}+p_{\mu}^{2}-2p_{\pi}p_{\mu}cos\phi+m_{\nu}^{2}}
\label{CQ}
\end{eqnarray}
from which one can obtain the square of the neutrino mass
\begin{eqnarray}
m_{\nu}^{2}&=&(E_{\pi}-E_{\mu})^{2}-(p_{\pi}^{2}+p_{\mu}^{2}-2p_{\pi}p_{\mu}cos\phi)= \nonumber \\
&=&(E_{\pi}-E_{\mu}-p_{\nu})(E_{\pi}-E_{\mu}+p_{\nu})= \nonumber \\
&=&m^{*}m^{**}
\label{M2}
\end{eqnarray}
where only the $m^{*}$ factor can approach zero and
where $m^{**}>>m^{*}$ and always positive. 
It is clear that
from relation~(\ref{M2}), given the masses of the pion and muon,
the momenta of the pion and muon, and the angle $\phi$, one can calculate
the square of the neutrino mass as a function of 
the above-mentioned parameters.

\subsection{On the upper limit of $m_{\nu}$}

A point or an interval estimator of an unknown parameter 
is any statistic whose value 
is a meaningful guess for the value of the unknown parameter which 
is assumed to have some fixed value~\cite{EDJRS}.
Sometimes experimental measurement may
yield non-physical values,
when the parameter value is near zero~\cite{FJMR}.
In these cases a statistical procedure can be used for 
estimation of the upper limit value of the parameter.
In the present analysis the neutrino mass
is the parameter to be estimated.
The upper limit of the mass of 
the muon neutrino~\cite{PDG} has a value near zero,
and is about seven orders of magnitude
smaller than the values of the measured experimental quantities
of the $\pi$$\rightarrow$$\mu$$\nu_{\mu}$ decay.
In determining such a mass in an unbiased approach one 
would necessarily expect a non-physical result for
$m_{\nu}^{2}$ to occur half of the time. This happens because
of the uncertainties of 
the experimental values of the physical quantities.
One should then make statistical 
inferences from observation
of the squared mass $m_{\nu}^{2}\pm\Delta$m$^{2}$ 
when $m_{\nu}^{2}$ is negative 
or near the non-physical region. $\Delta$m$^{2}$ is the standard deviation. 
It is possible to choose a ``classical'' confidence 
level {\it p} for the squared mass such that the corresponding classical 
confidence limit $m_{p,cl}^{2}$ is in the physical region.
When the variable $m^{2}$ has a Gaussian distribution 
the upper
limit is
\begin{equation}
m_{p,cl}^{2}=m^{2}+Z_{p}\Delta m^{2}
\label{otto}
\end{equation}
at a p$\%$ confidence level, 
where $Z_{85}$ = 1.036, $Z_{90}$ = 1.282, $Z_{95}$ = 1.645 
and  $Z_{97.5}$ = 1.960 (see ref.~\cite{CROW}). 
The classical confidence 
limit satisfies the probability statement
\begin{equation}
P(m_{o}^{2}<m_{p,cl}^{2})=p         .
\label{nove}
\end{equation}
This statement says that $m_{p,cl}^{2}$ has the probability 
{\it p} of being larger than the true value ($m_{o}^{2}$), 
whatever $m_{o}^{2}$ 
really is. It means that, if one repeats the experiment many times,
and each time one recalculates the value of  $m_{p,cl}^{2}$, 
then in $p\%$ of the cases $m_{p,cl}^{2}$ 
will be greater than the true value  $m_{o}^{2}$. 

\subsection{Monte Carlo events}

To obtain an interval estimate for the muon neutrino mass
from our measurement of the scattering angle and 
of the momenta of the two charged particles
of the decay, a statistical procedure has been applied. It
takes into account the measured pion and muon 
momenta and the angle
between them. These are distributed 
with Gaussian p.d.f.'s with means equal to the
measured values and standard deviations equal to 
the corresponding measured uncertainties.
A code has been written. Using the Monte Carlo method 
a set of $10^{5}$ decays has been 
generated. Each element of the set has been obtained extracting 
at random the momenta of the pion and of the
muon, and the angle between  the two particles
assuming the mass values of $m_{\pi}$ = (139.57018$\pm$0.00035) MeV
and $m_{\mu}$ = (105.658369$\pm$0.000009) MeV,
as given in ref.~\cite{PDG}.
The three variables were extracted independently of each other.
Other relevant quantities were
then calculated.
The distributions of $m_{\nu}^{2}$,
 of $m^{*}$ and of $m^{**}$
in relation (\ref{M2}) have been obtained.
%Figure~\ref{Figure 3.} shows: (a)
%the distribution of the pion momentum;
%(b) the distribution of the muon momentum; and
%(c) the distribution of the angle between the decaying pion
%and the muon. The calculated mean values and variances
%correspond to the measured values, showing the
%correctness of the procedure. One should note that the 
%individual negative values of the pion momentum
%in Figure~\ref{Figure 3.}(a) have no physical significance
%on their own, since we are only dealing with random distributions.
%Physical significance, here, is only to be attributed to
%the parameters (mean value and standard deviation) of
%the distributions used.
%\begin{figure}
%\begin{center}
%\epsfig{file=neufig3.eps,height=15cm,width=10cm}
%\caption{ Pion (a) muon (b) momentum and angle between pion and
%muon (c) distributions for all the generated events. 
%The data have been best fitted
%(dashed line) with Gaussians with mean values and standard deviations
%given by $p_{\pi}$ = (0.05$\pm$0.10) MeV/c (a);
%$$p_{\mu}$ = (29.90$\pm$0.15) MeV/c (b);
%$\phi$ = (163.0$\pm$1.0) deg (c).}
%\label{Figure 3.}
%\end{center}
%\end{figure}

\subsection{Analysis of the results}

\begin{figure}
\begin{center}
\epsfig{file=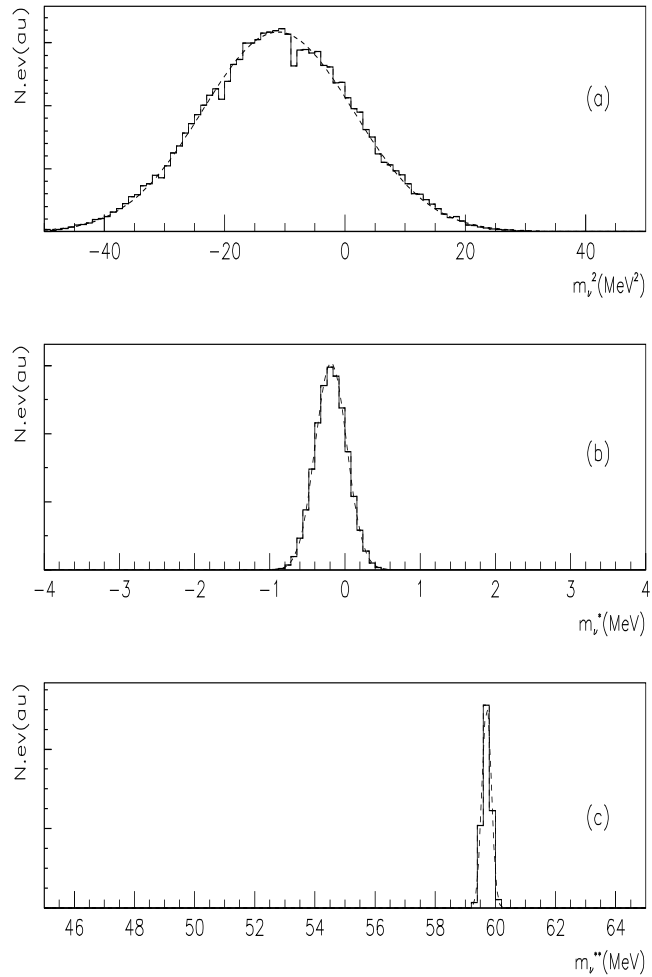,height=15cm,width=10cm}
\caption{ Distribution of $m_{\nu}^{2}$ (a), 
of $m^{*}$ (b), and of $m^{**}$ (c) generated with the
Monte Carlo method. Note the different scales along the horizontal
axes.
The data have been best fitted
(dashed line) with Gaussians with the following respective 
mean values and standard deviations: ($-$11.1; 12.5) MeV$^{2}$ (a);
($-$0.186; 0.211) MeV (b);
(59.7; 0.2) MeV (c).}
\label{Figure 4.}
\end{center}
\end{figure}

Figure~\ref{Figure 4.} shows the distributions of $m_{\nu}^{2}$
(upper part), $m^{*}$ (central part) and of $m^{**}$ (lower part).
The $m_{\nu}^{2}$ distribution is Gaussian, with
a negative mean value 
and relatively large standard deviation that depends
on the uncertainties in the experimentally
measured quantities. The  $m^{*}$ distribution is
Gaussian with a negative mean value    
and relatively small standard deviation.
Both distributions have ratios of the standard 
deviation to the mean value of the same order
of magnitude. 
The $m^{**}$ distribution is Gaussian but in the
physical region, with a mean value of 
(59.7$\pm$0.2) MeV, about two
orders of magnitude
greater than $m^{*}$.

The $m_{\nu}^{2}$ distribution leads to the squared
muon neutrino mass
\begin{equation}
m_{\nu}^{2}=(-11.1\pm12.5) \mbox{ MeV$^{2}$}
\label{nove}
\end{equation}
which is compatible with zero.
According to the classical method~\cite{FJMR}, described 
in Section 3.2, equation~(\ref{nove}) 
corresponds to the muon neutrino mass upper limit
\begin{equation}
m_{\nu}<\sqrt{-11.1+1.282\cdot12.5}\mbox{ MeV = 2.2 MeV}
\label{dieci}
\end{equation}
at the 90$\%$ confidence level. This is the lowest estimate
of the upper limit value of the muon 
neutrino mass
obtained with a visualizing detector, using all the directly
measured kinematic and dynamic parameters of the
$\pi$$\rightarrow$$\mu$$\nu_{\mu}$ decay, that is $p_{\pi}$,
$p_{\mu}$ and $\phi$ (see ref.~\cite{ROB} and~\cite{HYMA} and 
references quoted therein). 
The last measurement of this
kind of events has been performed with a helium
filled bubble chamber~\cite{HYMA}.
Because of the density of liquid helium, 
the muon tracks were only 1 cm long, and
composed by about (15$\pm$2) bubbles.
The authors assumed that all the stopped pions
decayed at rest.
For this reason they did not use in the analysis
the values of the angle
$\phi$ and of the momentum of each pion.
It is quite likely that most of the low 
energy pions decayed in flight.

Since then many 
experiments have been
devoted to measuring the muon momentum with higher accuracy.
Recently, Assamagan et al.~\cite{ASSA},
using a magnetic spectrometer equipped with a silicon microstrip
detector, studied the decay of pions stopped in a graphite
target. The authors measured for the muons an
average momentum of (29.79200$\pm$0.00011) MeV/c.
For the pions immediately before their decay, the estimated
average kinetic energy was T$_{\pi}$ = (0.425$\pm$0.016) eV,
so most of the pions actually decayed in flight.
Setting the energy of the pions and the angle $\phi$
to zero, the authors
deduced a squared muon neutrino mass of 
($-$0.016$\pm$0.023) MeV${^2}$ and,
according to the Bayesian approach, deduced the
corresponding
neutrino mass upper limit of m$_{\nu}$ $<$ 0.17 MeV (C.L. = 0.9).

Applying the procedure described in paragraph 3.2,
we have estimated the upper limit
of the muon neutrino mass from the data
of ref.~\cite{ASSA}, taking into account the measured
average energy value of the decayed pions and
assuming a non-zero angular uncertainty  $\Delta\phi$ of  0.2 deg
(differently from ref.~\cite{ASSA}).
In this case the distribution of $m_{\nu}^{2}$ is totally
in the physical region. The value obtained for the
muon neutrino mass is $m_{\nu}$ = (800$\pm$80) keV.

If we set the momentum of the decayed pion equal
to zero, as in ref.~\cite{ASSA}, if we assume
the uncertainty of this value to be of the same order
of magnitude as that measured for the muon, 
and take $\Delta$$\phi$ = 0.2 deg,
the distribution of $m_{\nu}^{2}$ extends
in the non-physical region. 
The calculated
upper limit of the muon neutrino mass is m$_{\nu}$ $<$ 110 keV
(C.L. = 0.9) using the classical approach~\cite{FJMR}.

These exercises reveal the necessity, 
for an accurate estimation
of the squared muon neutrino mass, of
precise and direct measurements of all kinematic
parameters of individual $\pi$$\rightarrow$$\mu$$\nu_{\mu}$
decay events, like those performed
for the event reproduced in Figure~\ref{Figure 2.}. 

Analysing the distributions 
shown in Figure~\ref{Figure 4.}, we studied the
characteristics of the statistical variables
$m_{\nu}^{2}$, $m^{*}$ and $m^{**}$.
Although both $m^{*}$ and $m^{**}$ have the same
Gaussian distribution (being composed of the same
physical quantities), the $m^{**}$ distribution
is shifted toward large positive values. The
$m^{*}$ distribution is close to zero and the 
$m^{*}$ values may be negative (this is true, also,
for the $m_{\nu}^{2}$ distribution).

Using the formulae of paragraph 3.1 we 
calculated the kinematic parameters
of the decay $\pi$$\rightarrow$$\mu$$\nu_{\mu}$,
assuming the mass values of $m_{\pi}$
 and $m_{\mu}$ as given
in paragraph 3.3, and 
a value for the muon neutrino mass of 0.300 keV.
The choice of this value for the neutrino mass is
related to the actual precision of the pion and 
muon masses~\cite{PDG}.
The mean value of the pion momentum was set to 0.05 MeV/c
and of the angle $\phi$ to 163.0 deg in order to allow a 
direct comparison with our measurements.
Applying the procedure described in paragraph 3.3
and running the Monte Carlo code with different values 
of the uncertainties in the input variables
(within 10$^{-10}$$\div$10 MeV/c for the momenta
$p_{\pi}$ and $p_{\mu}$, and 
 10$^{-5}$$\div$1 deg 
for the angle~$\phi$),
sets of 10$^{5}$ $\pi$$\rightarrow$$\mu$$\nu_{\mu}$
decays have been generated. Each decay has been obtained
extracting at random, independently, the three Gaussian
variables~$p_{\pi}$,~$p_{\mu}$ and $\phi$.
The procedure gives in output all the relevant
kinematic and dynamic quantities of the decay,
in particular, $m^{*}$, $m^{**}$ 
and $m_{\nu}^{2}$.

\begin{figure}
\begin{center}
\epsfig{file=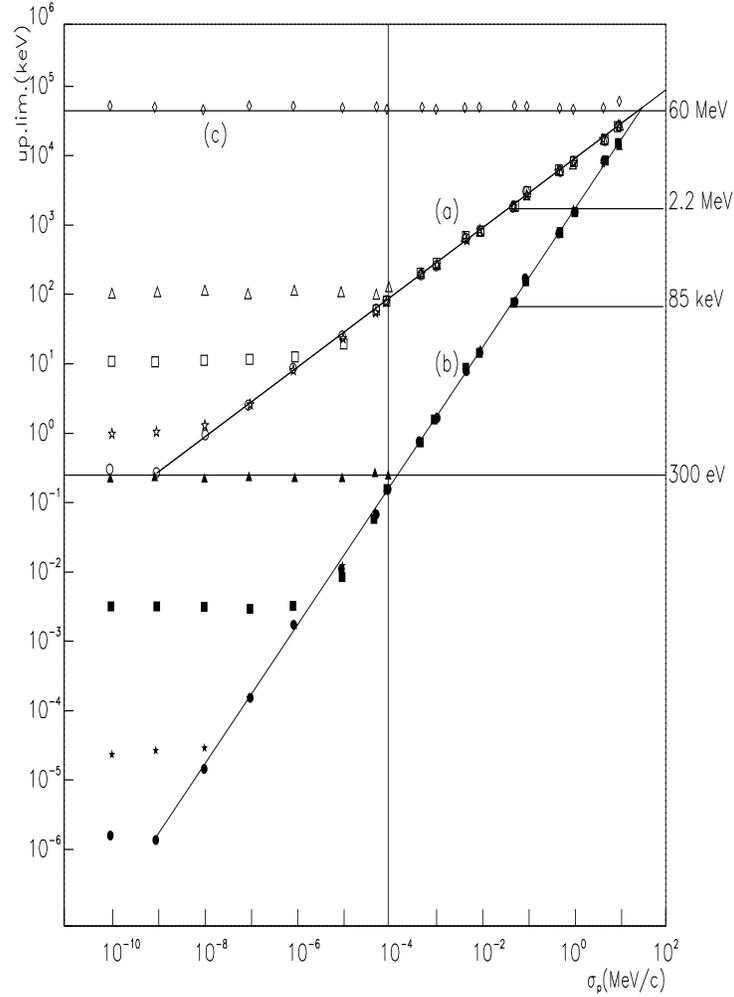,height=15cm,width=10cm}
\caption{ Behaviour of the  upper limits of $m_{\nu}$
(open triangles, squares, stars, circles), $m^{*}$ 
(full triangles, squares, stars, circles), 
and $m^{**}$ (open rhombs) calculated
from  the $m_{\nu}^{2}$, $m^{*}$ and $m^{**}$ distributions 
at 90$\%$ confidence level,
as a function of the experimental uncertainties 
in the momenta, and for
uncertainties in the angle $\phi$
of 1.0 deg (open triangles, full triangles), of 10$^{-2}$ deg 
(open squares, full squares), of 10$^{-4}$ deg (open stars, full stars) 
and of 10$^{-5}$ deg (open circles, full circles).
The lines (a) and (b) are drawn to guide the eye  for 
$\Delta\phi$=10$^{-5}$ deg and $\sigma_{p}>$10$^{-9}$ MeV/c.
The line (c) represents the upper limit value of 60 MeV.
The values of 2.2 MeV and 85 keV are 
from Table~\ref{Table II}.}
\label{Figure 5.}
\end{center}
\end{figure}

In Table~\ref{Table I} we present the results for different
sets of Monte Carlo events, all obtained  with an 
uncertainty of 1$^{o}$ in the measurement of the $\phi$ angle.
The choice of this value was again dictated by the uncertainty
of our measurement.
\begin{footnotesize}
\begin{table}[h]
\begin{center}
\begin{tabular}{|c|c||c|c||c|c||c|c|}
\hline
$\sigma_{\phi}$&$\sigma_{p_{\pi,\mu}}$&\multicolumn{2}{c||}{$m^{*}$}&\multicolumn{2}{c||}{$m^{**}$}&\multicolumn{2}{c|}{$m_{\nu}^{2}$}\\
(deg)&(MeV/c)&\multicolumn{2}{c||}{(MeV)}&\multicolumn{2}{c||}{(MeV)}&\multicolumn{2}{c|}{(MeV$^{2}$)}\\
\cline{3-8}
&&mean &standard &mean &standard &mean &standard \\
&&value&deviation&value&deviation&value&deviation \\
\hline
\hline
1.0&10&-4.3$\cdot$10$^{-1}$&15.0&59.9&11.0&-190.5&96$\cdot$10\\
1.0&1&-2.7$\cdot$10$^{-3}$&1.6&59.6&1.2&-1.9&93.0\\
1.0&10$^{-1}$&-1.0$\cdot$10$^{-4}$&16$\cdot$10$^{-2}$&59.6&1.2$\cdot$10$^{-1}$&-2.4$\cdot$10$^{-2}$&9.3\\
1.0&10$^{-2}$&-1.9$\cdot$10$^{-5}$&1.6$\cdot$10$^{-2}$&59.6&1.2$\cdot$10$^{-2}$&-1.3$\cdot$10$^{-3}$&0.93\\
1.0&10$^{-3}$&-1.3$\cdot$10$^{-5}$&1.6$\cdot$10$^{-3}$&59.6&1.2$\cdot$10$^{-3}$&-7.6$\cdot$10$^{-4}$&9.5$\cdot$10$^{-2}$\\
1.0&10$^{-4}$&-1.2$\cdot$10$^{-5}$&3.0$\cdot$10$^{-4}$&59.6&2.8$\cdot$10$^{-4}$&-7.2$\cdot$10$^{-4}$&1.8$\cdot$10$^{-2}$\\
1.0&10$^{-5}$&-1.2$\cdot$10$^{-5}$&2.5$\cdot$10$^{-4}$&59.6&2.5$\cdot$10$^{-4}$&-7.2$\cdot$10$^{-4}$&1.5$\cdot$10$^{-2}$\\
1.0&10$^{-6}$&-1.2$\cdot$10$^{-5}$&2.5$\cdot$10$^{-4}$&59.6&2.5$\cdot$10$^{-4}$&-7.2$\cdot$10$^{-4}$&1.5$\cdot$10$^{-2}$\\
\hline
\end{tabular}
\end{center}
\caption{Uncertainties, mean values, standard deviations of
the $m^{*}$, $m^{**}$ and $m_{\nu}^{2}$ distributions.
The data have been calculated assuming $m_{\nu}$ = 0.300 keV.
For momentum uncertainties less than 10$^{-6}$ MeV/c the values
remain constant, as shown in Figure~\ref{Figure 5.}.}
\label{Table I}
\end{table}
\end{footnotesize}
In the first two columns the uncertainties of the input
variables are reported. The third, the fourth and the fifth
columns show the mean values and the standard deviations
of the~$m^{*}$,~$m^{**}$ and $m_{\nu}^{2}$ distributions.
Both
$m^{*}$ and $m_{\nu}^{2}$ distributions 
have negative mean values.
The $m^{**}$ mean values are always positive. So,
for all the experimental uncertainties
reported in Table~\ref{Table I}, 
the distributions of $m^{*}$ and $m_{\nu}^{2}$ extend
into the non-physical region.
Using the same classical method applied to deduce
the muon neutrino upper limit mass from 
the $m_{\nu}^{2}$ distribution~\cite{FJMR},
we calculated an upper limit 
from the $m^{*}$ distribution.
Figure~\ref{Figure 5.} shows the
muon neutrino mass upper limits 
calculated from the $m_{\nu}^{2}$
distributions as a function of
the uncertainties of the momenta of the
$\pi$ and $\mu$ for
different values of the $\phi$ angle,
and for $m_{\nu}$ = 0.300 keV.
In the same Figure~\ref{Figure 5.}, the 
$m^{*}$ and $m^{**}$ (C.L. = 0.9)
upper limits are also reported. 
The values calculated from formula~(\ref{otto})
follow straight lines.
The lines (a) and (b) have been drawn to guide the eye,
the line (c) represents a constant upper limit value of 60 MeV. 
When the momentum
uncertainties are larger than about 10 MeV/c, the lines 
(a) and (b) intersect 
at the upper limit of $m_{\nu}$ $\approx$ 60 MeV. 
For momentum uncertainties less than 10$^{-9}$ MeV/c
and for angular uncertainties less than 10$^{-5}$ deg, 
the upper limit values of $m_{\nu}$
reach the assumed value of 0.300 keV, confirming 
the validity of the classical 
approach~\cite{FJMR} we used.
Moreover, one can see that the limit values of $m_{\nu}$ 
attainable with present modern
techniques correspond to measurement uncertainties
in momenta higher than 10$^{-4}$ MeV/c.

\begin{figure}
\begin{center}
\epsfig{file=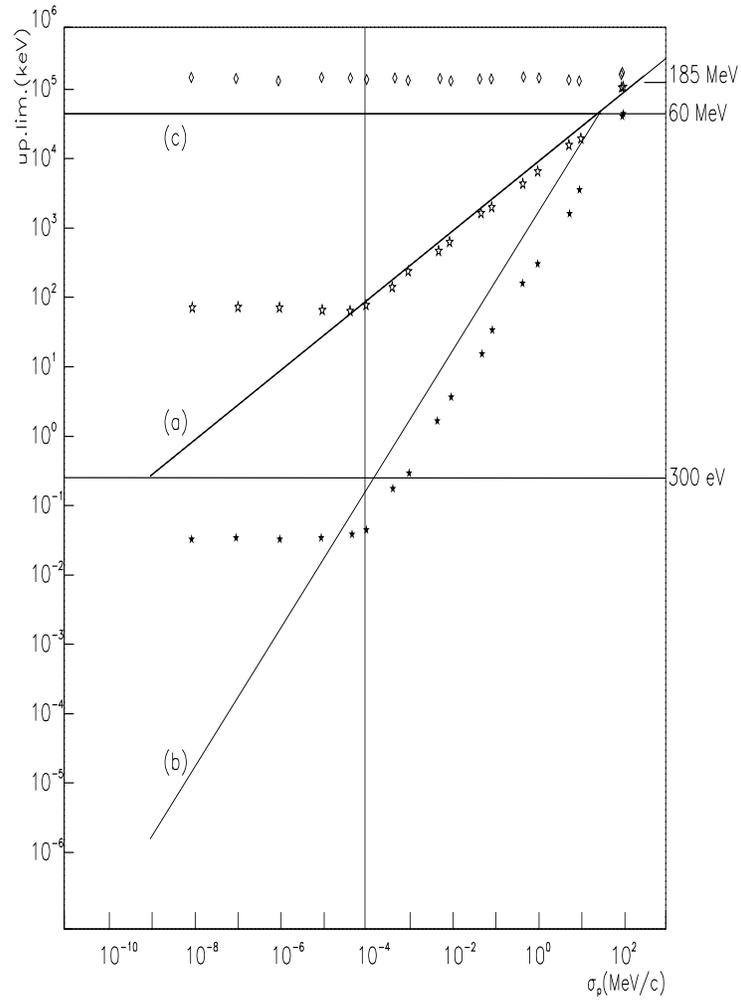,height=15cm,width=10cm}
\caption{ Behaviour of the  upper limits of $m_{\nu}$
(open stars), $m^{*}$ (full stars) and $m^{**}$ (open rhombs) 
calculated
from  the $m_{\nu}^{2}$, $m^{*}$ and $m^{**}$ distributions 
at 90$\%$ confidence level,
as a function of the experimental uncertainties 
in the momenta, and for
$\Delta\phi$ = 10$^{-4}$ deg, for pions
of 200 MeV/c. 
The lines (a) and (b) are those drawn to guide
the eye in Figure~\ref{Figure 5.}.
The line (c) represents the upper limit value of 60 MeV.}
\label{Figure 7.}
\end{center}
\end{figure}

To analyse the case
of pions decaying in flight with higher momenta,
we calculated the kinematic parameters of
the $\pi$$\rightarrow$$\mu$$\nu_{\mu}$ decay
for a pion momentum of 200 MeV/c
and neutrino mass of 0.300 keV.
This value of the pion momentum has been choosen
because, at the JINR Phasotron, the DUBTO experiment~\cite{DUBTO}
studying pion interactions
with $^{4}He$ nuclei at 200 MeV/c,
collected several $\pi$$\rightarrow$$\mu$$\nu_{\mu}$
events.
In this case the allowed values of the $\phi$ angle
are less than 11 deg. Setting the value of the
angle $\phi$ equal to 5 deg, we applied the
procedure described in paragraph 3.2.
The uncertainties of the momenta
were varied in the range 10$^{2}\div$10$^{-9}$ MeV/c and
of the angle $\phi$ in the range 1$\div$10$^{-5}$ deg.
Figure~\ref{Figure 7.} shows the 
same $m_{\nu}$, $m^{*}$ and $m^{**}$ upper limits
as in Figure~\ref{Figure 5.}, 
calculated using the $m_{\nu}^{2}$,
$m^{*}$ and $m^{**}$ distributions as functions of the
momentum uncertainties in the range 10$^{2}\div$10$^{-9}$ MeV/c,
and
with an error for the angle $\phi$ of 10$^{-4}$ deg.
As one can see,
at 200 MeV/c the $m_{\nu}$ upper 
limit values follow the
line (a) of Figure~\ref{Figure 5.}, 
which was drawn for
pions decaying at lower energies. 
The behaviours of the $m^{*}$ and $m^{**}$ upper limits
appear to be
nearly parallel to those
of Figure~\ref{Figure 5.}.

Considering the experimental uncertainties actually attainable
in laboratories and taking into account 
the present values of pion and muon
masses and their uncertainties,
it appears impossible to perform experiments, based
on the measurement of all the kinematic parameters of
the  $\pi$$\rightarrow$$\mu$$\nu_{\mu}$ decay, 
from which an upper
limit of the muon neutrino mass
less than about 1 keV can be deduced.

In the frame of the results of the
present analysis,
using the distributions of Figure~\ref{Figure 4.} 
relative to the measurement of the event displayed in
Figure~\ref{Figure 2.} and with the 
uncertainties of the present experiment,
we obtained the upper limits of $m_{\nu}$ and 
$m^{*}$, at given 
confidence levels, reported in Table~\ref{Table II}
using the classical statistical approach. 
The values have been deduced
with formulae~(\ref{otto}); in particular,
$m^{*}$ = ($-$0.186+1.282*0.211) MeV = 0.0845 MeV, 
at 90$\%$ confidence level. 
It must be noted that, with the upper limit 
of $m^{**}$ = 59.96 MeV,
one obtains
\begin{equation}
\sqrt{m^{*}m^{**}}=\sqrt{0.0845\cdot59.96}\mbox{ MeV}=\mbox{2.2 MeV}   .
\label{undici}
\end{equation}
The above values of $m^{*}$, $m^{**}$ and $m_{\nu}$ upper limits
are reported in Figure~\ref{Figure 5.}.
\begin{table}[h]
\begin{center}
\begin{tabular}{|c|c|c|c}
\hline
$m_{\nu}$(MeV)&$m^{*}$(keV)&p($\%$)\\
\hline
  1.4&33&80\\
  2.2&85&90\\
  3.1&161&95\\
\hline
\end{tabular}
\end{center}
\caption{Upper limits of the muon-neutrino mass (deduced
from $m_{\nu}^{2}$) and of $m^{*}$ at 
a p$\%$ confidence level,
for the event of Figure~\ref{Figure 2.} measured and analysed 
by the method described in the text.}
\label{Table II}
\end{table}

\section{Conclusions}

This paper describes the determination of
a confidence upper limit for the muon neutrino mass by measuring
the momenta of the decaying pion and of the produced muon
in a unique ($\pi\mu$e)
 nuclear event recorded in the PS 179 experiment
at LEAR of CERN. A sizeable improvement of previous results obtained 
using low density visualizing detectors
has been achieved.

\section{Acknowledgement}
We are grateful to S.~Bilenky, S.~Bottino,
L.~Cifarelli, A.~Kotzinian,
A.~Olchevski, N.~Russakovich for stimulating
discussions and useful comments, and to Miss V.~Rumiantseva for the
very accurate measurements with the microscope.

The Italian Ministry of Foreign
Affairs is acknowledged for its essential financial support to the
research.

\end{document}